\documentclass[apj]{emulateapj}

\usepackage{graphicx}
\usepackage{float}
\usepackage{subfigure}
\usepackage{verbatim}
\usepackage{color}

\usepackage{bm}

\def \araa{ARA\&A}
\def \rmp{Rev.~Mod.~Phys.}

\def \aap{A\&A}
\def \apjl{ApJL}
\def \apj{ApJ}

\def \mnras{MNRAS}
\def \pasj{PASJ}

\def \sci{Science}
\def \nat{Nature}

\newcommand*{\sgra}{Sgr\ A*}
\newcommand{\gb}{\gamma_{\rm j}\beta_{\rm j}}

\begin{document}

\title{Confronting the jet model of Sgr A* with the Faraday rotation measure observations}


\author
{Ya-Ping Li$^{1}$, Feng Yuan$^{2,1}$, Q. Daniel Wang$^{3}$}
\affil{$^{1}$Department of Astronomy and Institute of Theoretical Physics and Astrophysics,Xiamen University, Xiamen, Fujian 361005, China}
\affil{$^{2}$Shanghai Astronomical Observatory, Chinese Academy of Sciences, 80
Nandan Road, Shanghai 200030, China, fyuan@shao.ac.cn}
\affil{$^{3}$Department of Astronomy, University of Massachusetts, Amherst, MA 01003, USA, wqd@astro.umass.edu}

\begin{abstract}
{
\sgra\ is probably the supermassive black hole being investigated most extensively due to its proximity. Several theoretical models for its steady state emission have been proposed in the past two decades. Both the radiative-inefficient accretion flow and  the  jet model have been shown to  well explain the observed spectral energy distribution. Faraday rotation measure (RM) has been unambiguously measured at submillimeter wavelength, but has only been tested against the accretion flow model. Here we first calculate the RM based on the jet model and find that the predicted value is two orders of magnitude lower than the measured value. We then include an additional contribution from the accretion flow in front of the jet and show that the measured RM may be reconciled with the model under some tight constraints. The main constraint is that the inclination angle should be greater than $\sim 73^{\circ}$. But this requirement is not consistent with an existing observational estimate of the inclination angle. }
\end{abstract}

\section{Introduction}

Because of its proximity, \sgra\ at the center of our Galaxy presents us as a unique laboratory to study the low-luminosity accretion process of supermassive black holes (SMBHs; see Genzel, Eisenhauer \& Gillessen 2010; Falcke \& Markoff 2013; and \citealt{YuanNarayan2014} for the most recent reviews focusing on different aspects of \sgra). With a mass of $4\times10^6~M_\odot$ \citep{Sch02,Ghez08,Gillessen09,Meyer12},
\sgra\ has a bolometric luminosity of only $10^{36}~\textrm{erg s}^{-1}\approx3\times10^{-9}~L_{\rm Edd}$, the bulk of which is emitted in the  submillimeter bump. High spatial resolution \emph{Chandra} observations have resolved the Bondi radius, $R_{\rm Bondi} \approx 4^{\prime\prime}$( $\approx 4\times10^{5}R_{\rm S}$, where $R_{\rm S}\equiv2GM_{\rm BH}/c^2$ is the Schwarzschild radius of \sgra) given that the temperature of the ambient plasma as 1~keV (\citealt{Baganoff03}, Wang et al. 2013).
In combination with the inferred  gas density at the Bondi radius as $\simeq 100~\textrm{cm}^{-3}$\citep{Baganoff03}, the Bondi accretion rate is estimated to be $\dot{M}_{\rm B}\approx10^{-5}~M_\odot~\textrm{yr}^{-1}$, which is in good agreement with the three-dimensional numerical simulation for the accretion of stellar winds onto \sgra\ by \cite{Cuadra06}. The low luminosity combined with the Bondi accretion rate indicates that the radiative efficiency of \sgra\ is extremely low, $\eta\approx10^{-6}$.

In addition to the continuum spectrum, a high level of linear polarization ($2\%\sim10\%$) at frequencies higher than $\sim150$ GHz has been detected \citep{Aitken00,Bower05,Macquart06}.
This detection provides us with a tool to measure the magnetized plasma through the
Faraday rotation of the polarized light. The rotation of the polarization angle is expressed as $\chi=\chi_0+\rm{RM}\lambda^2$  (where $\chi_0$ is the intrinsic polarization angle, while $\lambda$ is the observed wavelength). The observed rotation measure (RM) of \sgra\ can thus serve as a probe of the electron density weighted by the line of sight components of the magnetic field toward these central engines\footnote{The first RM towards the extragalactic radio source was made by \cite{Cooper62}. Since then many VLBI observations (e.g., \citealt{Udomprasert97,Taylor98,Taylor00,Asada02,Zavala02,Zavala03,Gabuzda04}) have  been carried out to reveal the parsec-scale (i.e., jet) RM properties for quasars, radio galaxies, and BL Lac objects.  The time variabilities and distributions of the observed RMs suggest that they are primarily intrinsic to active galactic nuclei (AGNs).}. Obviously, any acceptable theoretical models must satisfy both the continuum spectrum and polarization observations.

\cite{Marrone07} made the first reliable determination of the RM for \sgra\ with simultaneous observations at multiple frequencies, using the Submillimeter Array (SMA) polarimeter; the mean RM is $-5.6\pm0.7\times10^{5}~\textrm{rad~m}^{-2}$ while a conservative 3 $\sigma$ upper limit  to the variation of the RM is  $2 \times 10^5~\textrm{rad~m}^{-2}$ \citep{Marrone07}. This unambiguous detection of the RM limits the accretion rate close to the black hole to be in the range of $(2\times10^{-7} - 2\times10^{-9})M_\odot~\textrm{yr}^{-1}$, depending on the configuration of the magnetic field.


Accompanied by the observational progresses, numerous theoretical efforts on modelling the quiescent state emission of \sgra\ have been made, including the development of the radiatively inefficient accretion flow (RIAF) model (\citealt{YQN03}; see \citealt{YuanNarayan2014} for the most recent review on RIAF theory and its applications), jet model (e.g., \citealt{FM00}, see also for the updated jet model by \citealt{Markoff07} and \citealt{Moscibrodzka13}), jet-advection dominated accretion flow (ADAF) model \citep{YMF02}
and spherical accretion model \citep[e.g.,][]{Melia92,Melia00}.

In particular, both the jet and RIAF models have successfully explained the spectrum \citep{FM00,YQN03} and the size of Sgr A* (\citealt{FM00,Yuan06,Huang07,Markoff07}, see \cite{Doeleman08} for the current inferred size from 1.3 mm VLBI observations). The two models are difficult to distinguish in this sense. The RIAF model has also explained the polarization and RM. The ``old'' version of the RIAF model (i.e., the classical ADAF without outflow, see \citealt{Narayan95,Manmoto97,Narayan98}) {\it assumes} that the accretion rate is  independent of radius. This leads to a large electron number density in the inner region of the accretion flow and thus a large RM, which significantly depolarizes the polarized emission.  Interestingly, before the detection of the radio polarization in Sgr A*, theoretical works had begun to show that very little mass available at large radii could actually accrete into the inner most region around the SMBH (e.g., \citealt{Stone99,Hawley02,Igumenshchev03}; see \citealt{Yuan12a} for a review). This development could then solve the polarization problem, as explained in the updated RIAF model presented in \cite{YQN03}. The inward decrease of mass accretion rate was explained either due to outflow \citep[ADIOS; e.g.,][]{Blandford99,Begelman12} or to convection \citep[e.g.,][]{Narayan00,Quataert00a}. Yuan, Bu \& Wu (2012; see also Narayan et al. 2012 and Li, Ostriker \& Sunyaev 2013) has shown that outflow rather than convection is responsible for the inward decrease of the accretion rate.  Such a theoretical prediction has been confirmed by the 3M second {\it Chandra} observations of Sgr A* \citep{Wang13}.

In this work, we calculate the RM for the jet model. Our calculations give RM far below the observed value mentioned above \citep{Marrone07}. Along the line of sight (LOS) toward the jet ``nozzle", the accretion flow should also contribute to the observed RM. This contribution depends on the viewing path through the structure of the accretion flow. We can thus either rule out the jet model or set a tight constraint on the viewing geometry.

In the rest of the paper, we present our RM calculations based on the jet model in Section 2 and calculate the RM contribution from the accretion flow in Section 3. Discussions on the implications of the results are given in Section 4. We summarize our results and conclusions in Section 5. Throughout this work, we assume the SMBH mass as $4\times10^6~M_\odot$ for \sgra\ and its distance to be 8 kpc.

\section{Jet model and its rotation measure}

We follow the jet model presented in \citealt{FM00} (hereafter FM00, namely the ``old" jet model). In this model the submillimeter emission of Sgr A* comes from the acceleration region of the jet close to the black hole, called ``nozzle''. The sharp spectral cut-off on the infrared (IR) side of the submillimeter bump requires a narrow electron energy distribution in the nozzle. Thus two cases are considered, namely 1) a power-law electron energy distribution with the index $q=2$ and with a steep cut-off at $5\gamma_{\textrm{e},0}$ and 2) a relativistic Maxwellian distribution with $\gamma_{\textrm{e},0}\approx\frac{kT}{m_\textrm{e}c^2}$. The ``free'' parameters in the nozzle component such as the magnetic field $B_0$, the electron Lorentz factor $\gamma_{\rm{e},0}$, the total electron density $n_0$, and the nozzle length scale $z_0$ (sonic point)  are constrained by the submillmeter spectrum. Beyond the jet nozzle,  the magnetized, relativistic plasma is ejected from the nozzle to form the extended jet, where the plasma is accelerated along the jet axis through the pressure gradient and expands sideways with its initial sound speed in units of the speed of light, $\beta_{\rm{s,0}}=\sqrt{\Gamma(\Gamma-1)/(\Gamma+1)}$ (where $\Gamma=4/3$ is the adiabatic index). The velocity field $\beta_{\rm{j}}$ in the extended jet with the bulk Lorentz factor $\gamma_{\rm{j}}$ follows the modified, relativistic Euler equation.  Algebraic transformations then lead to

\begin{equation}\label{euler}
{\partial\gb\over\partial z}
\left(\frac{\left(\frac{\Gamma+\xi}{\Gamma-1}\right)(\gb)^2-\Gamma}{\gb}\right)=\frac{2}{z}
\end{equation}
with
$\xi=\left(\gb/(\Gamma(\Gamma-1)/(\Gamma+1))\right)^{1-\Gamma}$.

\begin{figure*}
\begin{center}
\includegraphics[width=0.65\textwidth]{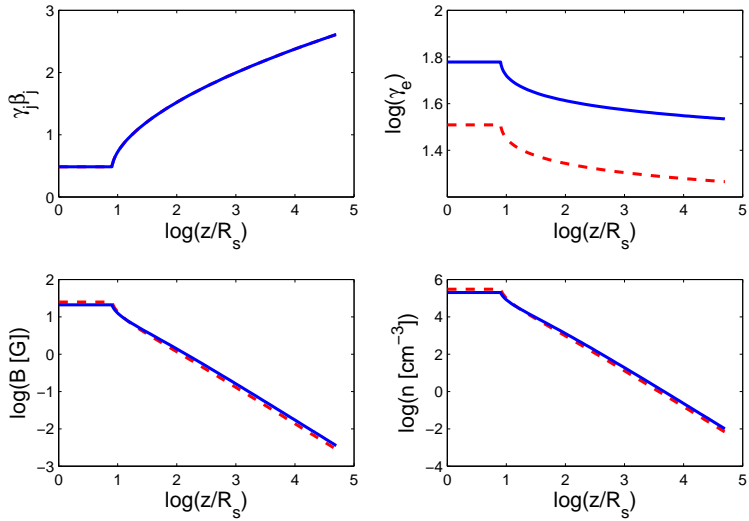}
\end{center}
\caption{\small The jet dynamical profile of $\gb$, Lorentz factor $\gamma_{\rm{e}}$, magnetic field $B$, and electron density $n_{\rm{e}}$ along the $z$-axis.  The solid (dashed) line is for the case of the power-law (Maxwellian) distribution of electrons.
\label{profile}}
\end{figure*}

The equation can be solved with the critical point condition $\beta_{\rm{j}}=\beta_{\rm{s,0}}$ at $z=z_0$. Taking adiabatic cooling due to the longitudinal pressure gradient and lateral expansion (FM00) into account, one can also determine the evolution profiles of the electron number density $n(z)$, magnetic field $B(z)$ and electron Lorentz factor $\gamma_\textrm{\small e}(z)$ along the $z$ axis in a cylindrical coordinate system, which are shown in Fig.~\ref{profile}.

\begin{figure}
\centering
\begin{minipage}{.45\textwidth}
\centering
\includegraphics[width=.8\textwidth]{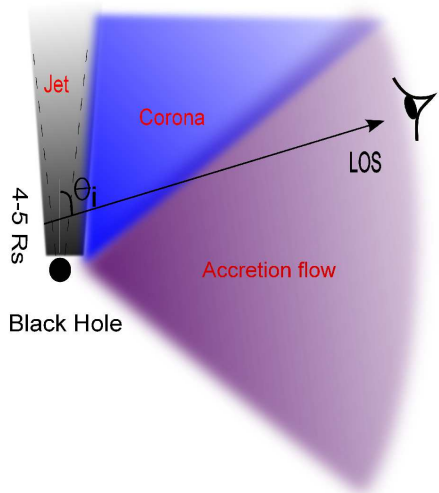}
\caption{\small The cross-section structure of the jet and accretion flow (including the corona) of Sgr A* in the $r-\theta$ plane (refer to Fig. 4 in Yuan \& Narayan (2014) for the more precise MHD numerical simulation result).  The black thick solid line denotes one representative sight-line through the plane. The dashed line presents the radial geometry of the magnetic field.
\label{illustration}}
\end{minipage}
\end{figure}

\begin{figure}
\begin{center}
\includegraphics[width=0.45\textwidth]{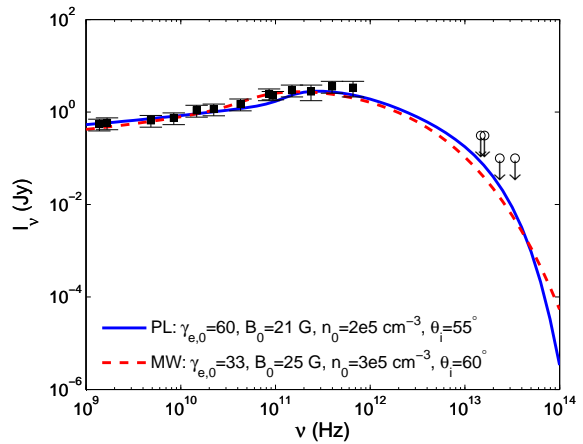}
\end{center}
\caption{\small The fitting spectrum of \sgra\ with the observation data from FM00 (infrared data shown as open circle are all upper limit.).  Although these data may not be the latest, we do not expect that any update would significantly affect our fitting and hence the following RM results. The solid blue line is for a power-law (PL) electron energy distribution with an index of $q=2$ and the dashed red line for a relativistic Maxwellian distribution (MW) with $\gamma_\textrm{e} = {kT}/{m_\textrm{e}c^2}$. The radius of the jet nozzle $r_0\simeq2.4 R_\textrm{s}$ and its height $z_0=8 R_\textrm{s}$ are assumed.
\label{jetfit}}
\end{figure}

The total jet synchrotron emission can be obtained by integrating along the $z$ axis. As we only care about the submillmeter emission, SSC process which only contributes to the high energy band (here X-ray) is neglected in the following calculation. The modification due to general relativistic effects on the spectrum is not included. Due to the Doppler boosting effect, we only consider the emission from the approaching jet.
The fitting spectrum with the nozzle parameters is shown in Fig~\ref{jetfit}.
The parameters we adopt to fit the spectrum are quite similar to those in FM00 (see also \citealt{Markoff07}), with only slight adjustments due to the different mass and distance of the SMBH adopted here. Specifically, the inclination angle $\theta_{\rm i}$ is determined in the spectral modelling via the Doppler boosting factor. Note that the difference of $\gamma_{\rm{e,0}}$ for the Maxwellian electron energy distribution with FM00 is only because of the different definition of $\gamma_{\rm{e,0}}$.

We then calculate the RM in the nozzle region along our LOS as
\begin{equation}\label{eq:rm}
  \textrm{RM} = 8.1\times10^5\int \frac{\log\gamma_e(z)}{2\gamma_e^2(z)}n_e(z)\textbf{\emph{B}}\cdot{\textbf{\emph{dl}}}~ \textrm{rad~m}^{-2}
\end{equation}
for the electron density $n_e$ in units of cm$^{-3}$, the path length \textbf{\emph{dl}} in units of pc, and the magnetic field \textbf{\emph{B}} in units of Gauss. The factor $\log\gamma_e(z)/2\gamma_e^2(z)$ is due to the relativistic correction \citep{Quataert00}. In the thermal case, the electron temperature $T$ is related to the Lorentz factor $\gamma_{\rm{e}}$ by $\gamma_e=kT/m_{e}c^2$. The above equation should be applied to the external Faraday rotation, which means that the emitting region and Faraday rotator exist separately. For the case of internal Faraday rotation considered here, namely, both the emission and Faraday rotator are contributed by the jet, there is an additional correction factor of $\sim1/2$ to the Eq.~\ref{eq:rm} \citep{Burn66,Homan12}\footnote{The correction factor for internal Faraday rotation depends on the physical and viewing geometry of the emission region. The factor of 1/2 is only true for an cylinder whose axis is parallel to the LOS and $\Delta\chi<90^{\circ}$. For other case, this factor differs a little but is also around 1/2 \citep{Cioffi80}.} .  Following the GRMHD numerical simulations of the accretion flow and jet formation (see review by Yuan \& Narayan 2014 and Fig. 4 of that paper), we find that the magnetic field in the jet is mainly radial and contains no reversals along our line of sight. 
 The ordered magnetic field we adopt presents an upper limit of the RM considered here.

In the case of thermal electrons, the integration of Eq.~\ref{eq:rm} for the thermal electrons in the nozzle region gives

\begin{equation}
\rm{RM} \approx 7.7\times10^3~\textrm{rad~m}^{-2},
\end{equation}
which is two orders of magnitude less than the observed value \citep{Marrone07}.  In the case of power-law electrons, the result is similar, which is $\textrm{RM}\approx 1.2\times10^3~\textrm{rad~m}^{-2}$. The point of the integration begins from the far side of the jet surface at the $4-5R_s$ from the black hole (see Fig.~\ref{illustration}) to the near side of the jet. This integration all the way down to the innermost region of the jet nozzle neglects the optical depth effect on its RM. If the emitting plasma is optically thick, this calculation should be regarded as an upper limit, which means that the corresponding RM will be far below the the observed value. We feel that there is little room to adjust the model parameters to increase the RM value to match the observed one. Compared to the RIAF model of \cite{YQN03}, the strength of the magnetic field is similar,  the small value of the RM is mainly because of the electrons density in the nozzle being much lower than that in the accretion flow. The dramatic discrepancy between the jet and the observation for the RM thus puts a challenge to the jet model.

\section{Accretion flow contribution to the RM }



A possible contribution to the observed RM, in addition to the jet itself, may come from the accretion flow around \sgra\ (see the illustration shown in Fig.~\ref{illustration}).  
The RM contribution from the accretion flow depends on the distributions of the electron density and magnetic field in the  $r-\theta$ plane of the accretion flow. We obtain such information from the multidimensional MHD numerical simulations (\citealt{DeVilliers05,Yuan12a}).

Based on a numerical simulation with the radial dynamical range as large as four orders of magnitude, Yuan, Wu \& Bu (2012; see also references therein) show that the radial density profile of the accretion flow can be well described by a power-law form, namely $n_{e}\propto{r^{-p}}$, with the index $p$ in the range of $0.5-1.0$. The temperature profile is described by $T_{\rm{e}}\propto{r^{-1}}$.

The strength of the magnetic field is tied to the density roughly by a parameter $\beta~(\equiv p_{\rm{gas}}/(p_{\rm{gas}}+p_{\rm{mag}})$, where $p_{\rm{gas}}$ and $p_{\rm{mag}}$ are the gas and magnetic pressures). The value of $\beta$ is typically $\sim 0.9$ (see review by Yuan \& Narayan 2014). But for the ``magnetically arrested disk'' (``MAD'') \citep{Narayan03,Tchekhovskoy11,Tchekhovskoy12,Narayan12,McKinney12}, the value of $\beta$ can be smaller, i.e, the magnetic field is stronger. But even in this case, current MHD simulations indicate that the field energy can at most be in equipartition with the thermal energy, i.e., $\beta\ga 0.5$. We therefore consider $0.5\la\beta\la0.9$. The configuration of the magnetic field is less well determined. The MHD numerical simulation can give us some suggestions, which, however, somewhat depend on the initial configuration of the field.
Without losing generality, we can simply assume a radial configuration. In reality, this should be multiplied by a factor of ``$\cos(\alpha)$'' to account for the effect of the inclination angle $\alpha$ of the magnetic field relative to our LOS. This would greatly reduce the RM contribution from the accretion flow. So our calculation for the case of pure radial magnetic field gives an upper limit to the RM contributed by the accretion flow, which we will see will not change our conclusion presented below.
Furthermore, based on  fully relativistic three-dimensional numerical simulations of accretion flows under the Kerr metric, \cite{DeVilliers05} find that the density and gas pressure decrease exponentially with polar angle (see Fig. 3 of that paper).

\begin{figure}
\begin{center}
\includegraphics[width=0.45\textwidth]{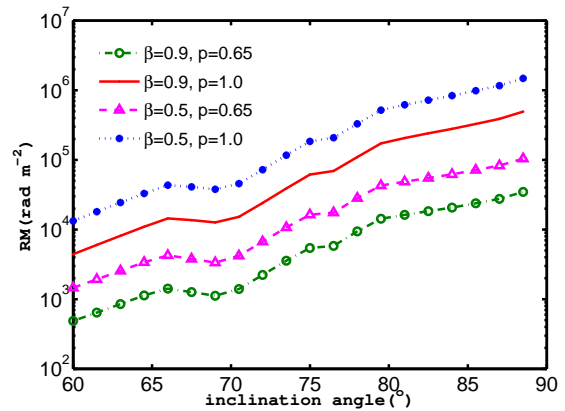}
\end{center}
\caption{\small RM contribution from the accretion flow as a function of the viewing angle for four different model parameter sets ($p$, $\beta$).  Only the red solid and blue dotted lines, which correspond to $p=1.0$ and $\beta=0.5-0.9$, can reach the magnitude of the observed RM, $\sim10^5~\textrm{rad~m}^{-2}$.
\label{RMacc}}
\end{figure}

With the corresponding distributions normalized to the values at the Bondi radius as inferred from {\it Chandra} data \citep{Baganoff03,Wang13}, we obtain the density, temperature, and magnetic field at each point in the $r-\theta$ plane of the accretion flow. The integration of Eq.~\ref{eq:rm} along the line of sight then gives the corresponding RM as a function of the viewing angle for different  parameters $p$ and $\beta$.  The jet inclination angle ($\theta_{\rm i}$) is also considered as a free parameter, although it has a preferred value (see later discussion).  Fig.~\ref{RMacc} shows that a smaller $\beta$ (i.e., stronger magnetic field) and/or a larger $p$ (i.e., weaker wind) can produce a larger RM. This is easy to understand. Only for the red solid and blue dotted lines, which correspond to $p=1.0$ and $\beta=0.5-0.9$, plus an inclination angle  $\theta_{\rm i}\ga 73^{\circ}$, or equivalently the angle between our line of sight and the equatorial plane of the accretion flow $\la 17^{\circ}$, can we obtain the magnitude of the observed RM, $\sim10^5~\textrm{rad~m}^{-2}$.

Whether the above values of $p$ and $\beta$ are reasonable? As stated above, the MHD numerical simulations of the hot accretion flow have shown that $\beta=0.5-0.9$ is reasonable. The value of $p=1$ is close to its upper limit obtained from numerical simulations and the modelling of \sgra\ \citep{YQN03}, although it is larger than that obtained from fitting the iron lines, which is $p\sim0.5$ \citep{Wang13}. Given the model and measurement uncertainties, however, we can conclude that the above-mentioned ``required'' values of $p$ and $\beta$ are roughly consistent with the current theoretical and observational constraints. In this case, the main constraint will be on the inclination angle. The required large angle means that  our LOS should intersect with the accretion disk close to its equatorial plane.

\section{discussion}

Is this inclination angle requirement consistent with the reality?
First, there are a few model constraints on the orientation of the black hole spin. But all these constraints are based on the assumption that the observed submillimeter radiation of Sgr A* comes from the hot accretion flow rather than the jet, in either the analytical RIAF model (\citealt{YQN03}; see also \citealt{Huang07,Broderick11}) or the MHD numerical simulations (e.g., \citealt{Sharma07,Moscibrodzka09,Dexter10,Shcherbakov12}). Thus such constraints do not apply here.

Second, it is widely accepted that accretion materials onto \sgra\  are mainly supplied by winds from circumnuclear massive stars (e.g., \citealt{Cuadra08}). These stars are primarily in a disk, which has an inclination of $\theta_{\rm{i}}\sim 53^\circ$ and a line-of-nodes position angle of $\sim 100^\circ$ (east from north, \cite{Paumard06}). One may expect that the angular momentum of this stellar disk and its inclination are reflected in the accretion flow morphology  projected in the sky. Indeed, this expectation has been confirmed by both the X-ray observations of the quiescent X-ray emission from \sgra\  \citep{Wang13} and the recent Event Horizon Telescope observations at 1.3 mm (Psaltis et al. 2014).  The elongation direction is also consistent with the well-constrained value, $\sim 90-100^\circ$, inferred from the recent mm-VLBA observations \citep{Bower14b}. While the X-ray emission traces the accretion flow in outer regions, the mm emission in innermost ones. Therefore, we may consider that the inclination angle of the accretion flow is well determined by that of the stellar disk. Adopting the value of $53^\circ$ gives a RM less than $10^4~\textrm{rad~m}^{-2}$, much smaller than the observed value.


Finally, the jet model itself has a constraint on the inclination angle, which is $\theta_{\rm i}\approx 55^{\circ}$ or $60^{\circ}$ (for the ``old" jet model, but see below for the updated one) for the power-law or the thermal distributions, respectively. Either of these two values is smaller than the ``required'' value of $\theta_{\rm i}\ga 73^{\circ}$. This means that, if we reasonably assume that jet is perpendicular to the accretion flow, the jet model fails to explain the observed RM.

However, with the further consideration of the 7-mm VLBA data in the spectral energy distribution analysis, the updated jet model \citep{Markoff07} constrains the inclination angle as $\theta_{\rm{i}}\gtrsim75^\circ$ and a position angle in the sky as $105^\circ$.   With this large angle, we find that the updated jet model could have a RM chiefly in the foreground accretion flow comparable to the observed one. But in this case, the direction of the jet is completely inconsistent with that of the stellar disk.

Are there any other contributions to the RM in addition to the plasma within the jet and the accretion flow considered above? One possibility is from the ISM (interstellar medium).  However, several studies have concluded that the ISM contribution to the RM is much less than $\sim10^4~\textrm{rad~m}^{-2}$ due to the weaker magnetic field ($\lesssim$1 mG; e.g., \citealt{Han99,Baganoff03,Marrone07}). The recent discovered magnetar PSR J$1745-2900$ close to \sgra\ with a projected offset of $\sim 3^{\prime\prime}=0.12$ pc {shows a $\rm{RM}=(-6.696\pm0.005)\times10^4~\rm{rad~m^{-2}}$} \citep{Eatough13}.  The recent VLBA observations further demonstrate that PSR J$1745-2900$ shares the same scattering medium with \sgra\ \citep{Bower14a}. Therefore, one may conclude that the ISM contribution to the RM is far below the value observed toward \sgra\ itself \citep{Marrone07}.

Therefore, the ``old" jet model may be ruled out, whereas the new one requires that the jet direction deviates significantly from being  perpendicular to the accretion flow due to some unknown reasons.

\section{summary}

We have calculated the expected RM from the jet model of Sgr A*. We find that the contribution to the RM from the jet is estimated to be less
than $7.7 \times 10^3~\rm{rad~m^{-2}}$, while the observed RM is $5.6\pm 0.7 \times 10^5~\rm{rad~m^{-2}}$ (the minus sign ignored). We have then further considered the RM contribution from the sight-line passage through the foreground accretion flow and find that its angle relative to our LOS needs to be larger than $\theta_{\rm i}\ga 73^{\circ}$, assuming that the jet is perpendicular to the accretion flow. But this requirement of the inclination angle is inconsistent with either, $\theta_{\rm i} \approx 55^\circ - 60^\circ$, set by the ``old" jet model itself \citep{FM00} or $\sim53^\circ$, set by the stellar disk inclination. Therefore, the ``old" jet model can firmly be rejected. The updated jet model with a highly inclined LOS, however, cannot be ruled out by the current RM observation alone.  The jet direction predicted  by the model appears to be inconsistent with  the axis of the accretion flow. The future high precision imaging data lending more stringent constraints on the putative jet and/or the inclination and size of the accretion flow could finally break the degeneracy of the theoretical models.


\acknowledgments

{We thank Sera Markoff, Heino Falcke, Geoffrey Bower for the constructive discussions  and the anonymous referee for useful comments. This work was supported in part by the Natural Science Foundation of China (grants 11103061, 11133005, 11121062, and 11103059), the National Basic Research Program of China (973 Program, grant 2014CB845800), and the Strategic Priority Research Program ``The Emergence of Cosmological Structures" of the Chinese Academy of Sciences (grant XDB09000000).
}

\end{document}